\documentclass[preprint,aps,prb,longbibliography]{revtex4-2}

\usepackage{xcolor}
\usepackage{amsmath,amssymb}
\usepackage{comment}
\usepackage{siunitx}
\usepackage{here}
\usepackage{graphicx}
\usepackage{dcolumn}
\usepackage{bm}

\begin{document}


\title{Spatiotemporal visualization of a surface acoustic wave coupled to magnons across a submillimeter-long sample by pulsed laser interferometry}

\author{Kazuki Maezawa}
\author{Shun Fujii}%
\author{Kazuto Yamanoi}%
\author{Yukio Nozaki}%
\author{Shinichi Watanabe}%
\email{watanabe@phys.keio.ac.jp}
\affiliation{%
 Department of Physics, Faculty of Science and Technology, Keio University, 3-14-1 Hiyoshi, Kohoku-ku, 
Yokohama, Kanagawa 223-8522, Japan
}%

\date{\today}

\begin{abstract}
    Surface acoustic waves (SAWs) coupled to magnons have attracted much attention because they allow for the long-range transport of magnetic information which cannot be achieved by magnon alone. We employed pulsed laser interferometry to visualize the entire spatiotemporal dynamics of a SAW that travels  on a nickel (Ni) thin film and is coupled to magnons. It was possible to trace the coupling-induced  amplitude reduction and phase shift that occurs as the SAW propagates over a distance of 0.4 mm. The observed changes are consistent with results obtained from conventional radio-frequency transmission measurements, which probe the total SAW absorption due to magnon--phonon coupling. This result verifies that our method can accurately capture the spatiotemporal dynamics of a SAW coupled to magnons across the entire length of the sample. Additionally, we validated our time-resolved profiles by comparing them with theoretical results that take the echo wave due to reflection into account. The impact of the echo wave is significant even when it has propagated over a distance on the order of millimeters. Our imaging results highlight the visualization of the long-range propagation of the SAW coupled to magnons and offer more information about the surface vibration profiles in such devices.

\end{abstract}

\maketitle

\newpage

\section{\label{introduction}INTRODUCTION}
The generation of propagating elastic strain in solids serves as one possible route to generate magnons through magnon--phonon coupling \cite{Kittel1958, Akhiezer1959}. In particular, the coupling in the case of a surface acoustic wave (SAW) that propagates on a magnetic material has attracted significant interest. SAW-induced magnetoelastic phenomena were initially applied to  isolators \cite{Lewis1972} and phase shifters \cite{Ganguly1975}, and about a decade ago, Weiler {\it et al.} elucidated  the physics behind the ferromagnetic resonance (FMR) driven by magnon--phonon coupling \cite{Weiler2011}: to explain this phenomenon, they used a time- and space-dependent strain-induced tickle field that exists during SAW propagation, and they successfully predicted the dependence of the SAW attenuation on the  angle of the external magnetic field. Since then, numerous theoretical \cite{Dreher2012, Gowtham2016, Li2017, Verba2018, Verba2019, Yamamoto2020, Yamamoto2022} and experimental \cite{Gowtham2015, Labanowski2016, Xu2018, Bhuktare2019, Kurimune2020, Xu2020, Liu2021, Kawada2021, Tateno2021, Verba2021, Huang2023, Sasaki2021,Hatanaka2022, Hatanaka2023} studies concerning SAW-induced magnetoelastic phenomena have been reported. One interesting characteristic  of these phenomena is the rather long propagation distance: If a SAW couples to magnons, it facilitates long-range transport of magnetic information (over distances on the order of millimeters) \cite{Chen2017, Li2017, Casals2020}, which is usually difficult to achieve by using magnons induced solely by magnetic fields because magnons show rapid decay compared to SAWs due to magnetic losses \cite{Zhao2021, Chen2017}. In addition, generation  and detection of SAWs are  possible by using a pair of  interdigital transducers (IDTs), and this enables various experiments concerning SAW-induced magnetoelastic phenomena.

   Since SAW-induced magnon--phonon coupling is a phenomenon that depends on both time and space, direct spatiotemporal imaging is necessary. However, reports on the observation of the spatiotemporal dynamics of magnon--phonon coupling are rare;  most previous studies focused on transmission measurements of SAWs between a pair  of IDTs, and in such measurements, the spatial decay profile cannot be confirmed.  To obtain a more complete spatiotemporal profile, several techniques have been employed: spatially resolved photoemission electron microscopy to visualize strain waves \cite{Foerster2017, Foerster2019, Casals2020}, characterization of changes in the polarization direction of reflected light due to SAWs coupled to magnons  \cite{Kuszewski2018}, and the Brillouin light-scattering technique to directly visualize changes in the surface vibration amplitude induced by magnon--phonon coupling \cite{Zhao2021}. To the best of our knowledge, there are no reports on the quantitative evaluation of the amplitude- and phase-profiles of a SAW coupled to magnons across the entire length of a usual device. This prevents us from providing direct supporting evidence for the decay behavior of the SAW in such a system with magnon--phonon coupling. On the other hand, optical interferometry has been used  to measure the amplitude and phase of surface and bulk acoustic waves, and axial resolutions ranging from picometers to femtometers have shown to be possible \cite{Knuuttila2000, Kokkonen2008, Leirset2013, Iwasaki2022, Shao2022}. Such a resolution may also enable us to detect changes in surface waves caused by magnon--phonon coupling.
   
   In this study, we adopt pulsed laser interferometry with a stroboscopic optical sampling approach \cite{Shao2022} to record the spatiotemporal dynamics of a SAW coupled to magnons. We captured the spatial variations  in amplitude and phase of the SAW across an unprecedentedly long distance of 0.4 mm (across the entire Ni thin film). We also confirmed an interference effect due to the reflection of the SAW at the IDT used for detection, and we reproduced  the time-resolved profiles by a model that accounts for this reflection. The overall absorption of the SAW due to magnon--phonon coupling derived from the spatiotemporal profile is quantitatively consistent with the results from conventional radio-frequency (RF) transmission measurements. The spatiotemporal dynamics provide a better insight into the magnon--phonon coupling of this system, as we can confirm how the SAW propagates over distances on the order of millimeters.

\section{EXPERIMENTAL SETUP}
\subsection{Sample preparation}
\begin{figure*}[!t]
    \centering
    \includegraphics[width=\textwidth]{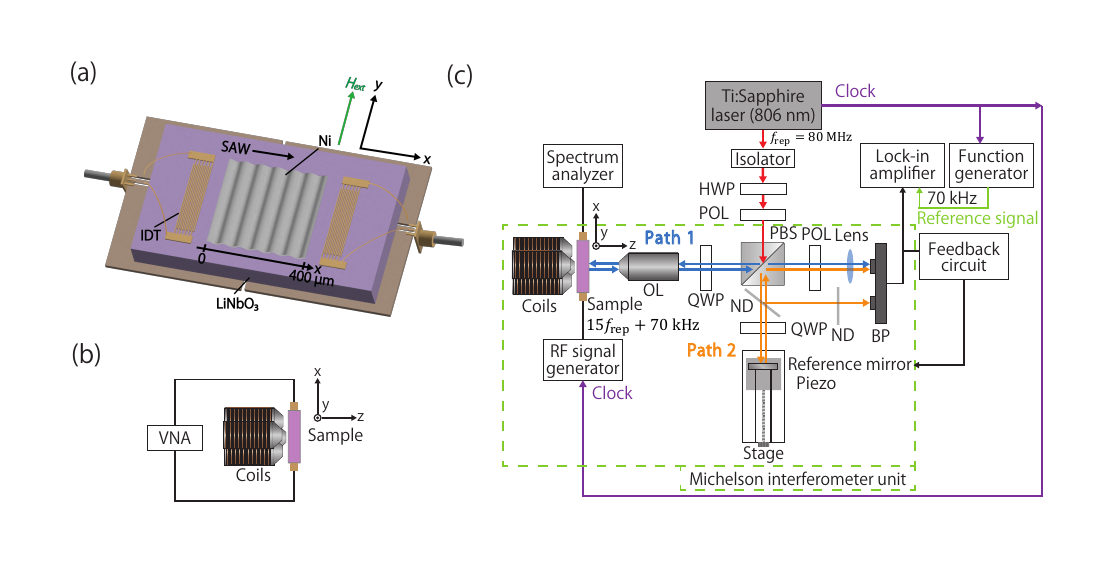}
    \caption{(a) Illustration of the sample structure consisting of two IDTs and a Ni thin film on a $\text{LiNbO}_3$ substrate. A SAW propagating on the Ni thin film along the $+x$ direction is also shown. An RF circuit board is used to connect the IDTs and coaxial connectors electrically. (b) Schematic of the experimental setup for the RF transmission measurements. Here, a vector network analyzer (VNA) is used to determine the transmission  parameter. (c) Schematic of the experimental setup used to perform pulsed laser interferometry. Stroboscopic optical sampling is implemented by using a SAW driving frequency that slightly differs from 15 $f_{rep}$. HWP: half-wave plate, POL: polarizer, PBS: polarizing beam splitter, QWP: quarter-wave plate, OL: objective lens (50X, numerical aperture $=0.55$), ND: variable neutral density filter, BP: balanced photodetector.}
     \label{sample}
\end{figure*} \par

    Figure \ref{sample}(a) depicts the sample structure. The piezoelectric material, a 128° \textit{Y}-cut $\text{LiNbO}_3$ substrate with a thickness of 0.5 mm, is shown in purple. The two IDTs for SAW generation and detection shown in brown were deposited onto the substrate by electron-beam (EB) evaporation. We used the following layer structure: titanium with a thickness of 3 nm and gold layer with a thickness of 70 nm. The spacing between the fingers of the IDTs is $\SI{3.75}{\micro m}$, and each IDT finger has a width of $\SI{3.75}{\micro m}$.  As shown in the figure, both IDTs are electrically connected to separate coaxial connectors. The SAW propagation direction is along the crystallographic $X$ axis of the $\text{LiNbO}_3$ crystal [the $x$ axis in Fig.~\ref{sample} (a)]. The fundamental resonant frequency of the SAW excitation  was 251 MHz according to our RF transmission measurements. The gray rectangular layer between the two IDTs illustrates the polycrystalline Ni thin film (during SAW excitation),  which was also deposited using EB evaporation. We used a film thickness of 50 nm, and the $x$- and $y$-extensions are $\SI{400}{\micro m}\times \SI{360}{\micro m}$, respectively.

\subsection{RF transmission measurements}

Figure \ref{sample}(b) shows a schematic of the experimental setup for the RF transmission measurements. The home-made  magnetic coils are located behind the sample. These coils allow for the application of magnetic fields to the sample in two orthogonal directions to change the strength of the magnon--phonon coupling  \cite{Weiler2011}.  To measure the frequency $\nu$  dependence of the transmission parameter \(S_{\mathrm{21}}\), the two coaxial connectors attached to the sample are connected to a vector network analyzer (VNA), and $S_{21}(\nu) $ is measured as a function of $H_{\text{ext}}$. This measurement was performed for various directions of $H_{\text{ext}}$. Furthermore, we applied time-domain gating to the obtained frequency-domain data $S_{21}(\nu)$  \cite{Archambeault2006} to isolate either the main response (the SAW that directly reaches the detection IDT) or the triple transit echo response [the SAW that is first reflected two times (one reflection per IDT) and then reaches the detection IDT] \cite{HashimotoBook}. This method effectively eliminates any spurious electromagnetic crosstalk. The Fourier transforms of the gated signals of the main response and the triple transit echo response were used to characterize the transmission properties of the SAW device.

\subsection{Pulsed laser interferometry with stroboscopic optical sampling}

The experimental setup used to perform pulsed laser interferometry with stroboscopic optical sampling is shown in Fig.~\ref{sample}(c). The excitation source is a mode-locked Ti:Sapphire laser with a repetition rate $f_{\text{rep}}$ of 80 MHz and a pulse width of about 100 fs. Each pulse of the emitted optical pulse train enters the Michelson-interferometer unit indicated by the green dashed rectangle. Here, the pulse is first divided into a pulse propagating along Path 1 (shown in blue) and a pulse propagating along Path 2 (shown in orange) by a polarizing beam splitter (PBS). The pulse propagating along Path 1 passes through an objective lens and is focused on the sample (the spot size was approximately \SI{1}{\micro m}), where it is reflected. Simultaneously, a SAW is propagating on the sample (we excited the SAW at $f_{\mathrm{SAW}} = 15f_{\text{rep}} + \SI{70}{kHz} \approx$ 1.2 GHz, by an RF signal generator with an input power of $+16$ dBm; this excitation frequency corresponds to the fifth harmonic of the fundamental resonance frequency of the SAW in our sample). Since the temporal width of the laser pulse is significantly shorter than the oscillation period of the SAW, the reflection of the laser pulse can be approximated by a reflection from a stationary sample surface. Moreover, since the spot size is smaller than the wavelength of the SAW (approximately \SI{3}{\micro m}), the laser pulse probes the local SAW-induced displacement of the sample surface. Regarding the relation between the probe timing and the surface displacement, because $f_{\mathrm{SAW}}$ is slightly faster than $15f_{\text{rep}}$, consecutive laser pulses do not probe the same surface displacement condition, but the same condition is probed  every 1/$|f_{\mathrm{SAW}} - 15f_{\text{rep}}| = 1/(\SI{70}{kHz})$. As a result, the interference signal between pulses coming from Path 1 and Path 2 is the same every 1/(\SI{70}{kHz}) and we obtain the modulated interference signal with a modulation frequency of $\SI{70}{kHz}$ at a balanced photodetector (BP). By analyzing the amplitude and phase of the \SI{70}{kHz}-frequency component  of the signals by a lock-in amplifier, we can determine the surface displacement at a given position $x$ on the surface. This is the stroboscopic effect used to investigate the spatiotemporal surface-vibration profile. 

By moving the sample on a motorized stage along the $x$-axis, we can obtain the spatiotemporal dynamics of the surface vibration.  The four coils were used to apply an external magnetic field $H_{\text{ext}}$ along the $y-$direction to change the strength of the magnon--phonon coupling. The temporal profile of the displacement in our experiment is determined as follows. First, we obtain the amplitude $R(x, \mu_0 H_{\text{ext}})$ and the phase $\theta^{\prime} (x, \mu_0 H_{\text{ext}})$ of the lock-in signal at the position $x$ and the external magnetic field $\mu_0 H_{\text{ext}}$. Then, we extract $\theta(x, \mu_0 H_{\text{ext}})$ by subtracting the phase at the Ni thin film edge $x=0$ at $\mu_0 H_{\text{ext}}=\SI{17.4}{mT}$ from $\theta^{\prime}(x, \mu_0 H_{\text{ext}})$, i.e., $\theta(x, \mu_0 H_{\text{ext}})= \theta^{\prime}(x, \mu_0 H_{\text{ext}}) - \theta^{\prime}(x=0, \SI{17.4}{mT})$. Finally,  the displacement at given $t$, $x$, and $H_{\text{ext}}$ is determined as $R(x, \mu_0 H_{\text{ext}})\cos{( 2 \pi f_{\mathrm{SAW}} t + \theta(x, \mu_0 H_{\text{ext}}))}$. The time origin is thus defined as the time instant when the surface displacement at $x=\SI{0}{\micro m}$ reaches its peak at $\SI{17.4}{mT}$. More details about the effect of the quarter-wave plate and the feedback control of the optical path length are provided  in Appendix A.

\section{Results}
\subsection{Magnetic field dependence of the RF transmission}
\begin{figure}[!t]
    \centering
    \includegraphics[width=0.7\columnwidth]{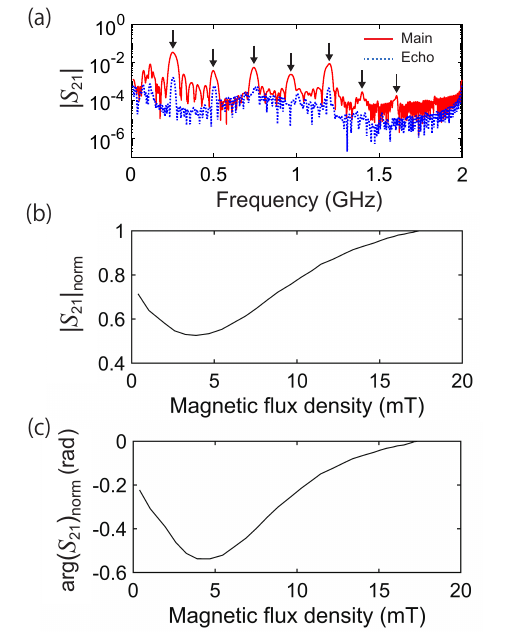}
    \caption{(a) Fourier amplitude spectrum of the SAW transmission, $|S_{21}(\nu)|$, of the main response (red solid curve) and the triple transit echo (blue dotted curve) for $\mu_0 H_{\text{ext}}=\SI{17.4}{mT}$ (the external magnetic field was applied along the $y$-axis).  The black arrows indicate the peak positions of the seven distinct peaks observed in these spectra. (b) Magnetic flux-density dependence of the amplitude of the main response at $\nu_0$ = 1.2 GHz normalized to the amplitude for a magnetic field of $\SI{17.4}{mT}$, $|S_{21} (\nu_0, \mu_0 H_{\text{ext}})|_{\text{norm}} = \frac{|S_{21}(\nu_0, \mu_0 H_{\text{ext}})|}{|S_{21}(\nu_0, \SI{17.4}{mT})|}$, and (c) that of the phase change of the main response at 1.2 GHz with respect to the phase for a magnetic field of $\SI{17.4}{mT}$, $\text{arg}(S_{21}(\nu_0, \mu_0 H_{\text{ext}}))_{\text{norm}} = \text{arg}(S_{21}(\nu_0, \mu_0 H_{\text{ext}})) - \text{arg}(S_{21}(\nu_0, \SI{17.4}{mT}))$.}
     \label{VNA}
\end{figure} \par

To characterize the magnon--phonon coupling in our sample using a well-established method, we initially performed RF transmission measurements for different values of the external magnetic field. First, we applied $H_{\text{ext}}$ in the $y$-direction and obtained the Fourier spectra $S_{21}(\nu)$ of the main response and the triple transit echo. Figure \ref{VNA}(a) shows the measurement results for $\mu_0 H_{\text{ext}}$ = 17.4 mT. Under this magnetic field, the magnon--phonon coupling in our sample is weak, because the FMR of the Ni thin film lies at significantly higher frequencies than the observed peaks. Seven distinct peaks can be identified in the transmittance spectrum of the main response, and the peak at the lowest frequency (251 MHz) corresponds to the fundamental frequency of the SAW. The other peaks correspond to the higher harmonics up to the seventh harmonic at 1.605 GHz. In the following experiments, we focus on the fifth peak at around 1.2 GHz. The so-called amplitude reflectivity at the IDT, $r_{\mathrm{IDT}}$, can be estimated from the ratio of $|S_{21}(\nu)|$ of the main response to that of the triple transit echo response, which is equal to the main response multiplied by $r_{\mathrm{IDT}}^2$. From the data, we estimated $r_{\mathrm{IDT}}$ = 0.19 at $\nu_0$ = 1.2 GHz.  

Figure \ref{VNA}(b) shows the magnetic flux-density dependence of the amplitude of $S_{21}(\nu_0)$ of the main response (the curve is normalized to the amplitude for $\mu_0 H_{\text{ext}}$ = 17.4 mT). The corresponding phase data (with respect to the phase for $\mu_0 H_{\text{ext}}$ = 17.4 mT) is shown in Fig.~\ref{VNA}(c). As the external magnetic flux density decreases, the amplitude becomes smaller and the phase becomes more negative; both reach their minimum values around $\mu_0 H_{\text{ext}} \approx$ 3.3 mT. This behavior is attributed to the fact that at this particular magnetic flux density, the FMR frequency of the Ni thin film matches the SAW excitation frequency of 1.2 GHz. This resonant excitation condition results in an efficient conversion of elastic energy to magnetic energy through magnon--phonon coupling \cite{Weiler2011}.

   In a separate experiment that is not shown here, we found that the reduction in the SAW amplitude due to magnon--phonon coupling is most pronounced if the external magnetic field is applied in the $y$ direction. It is known that the strain-induced RF magnetic field reaches its maximum in Ni if the angle between the magnetization direction and the $x$ direction is 45°, because here the gradient of the magnetic free energy is maximized \cite{Weiler2011}. Therefore, our separate experiment indicates that the magnetization direction is rotated by 45° with respect to the $x$ axis when the external magnetic field of $\mu_0 H_{\text{ext}} \sim$ 4 mT is applied in the $y$ direction. This alignment condition is a result of the minimization of the magnetic free energy, considering both the external magnetic field and the in-plane anisotropy field \cite{Hatanaka2022}.

\begin{figure*}[!t]
    \centering
    \includegraphics[width=\textwidth]{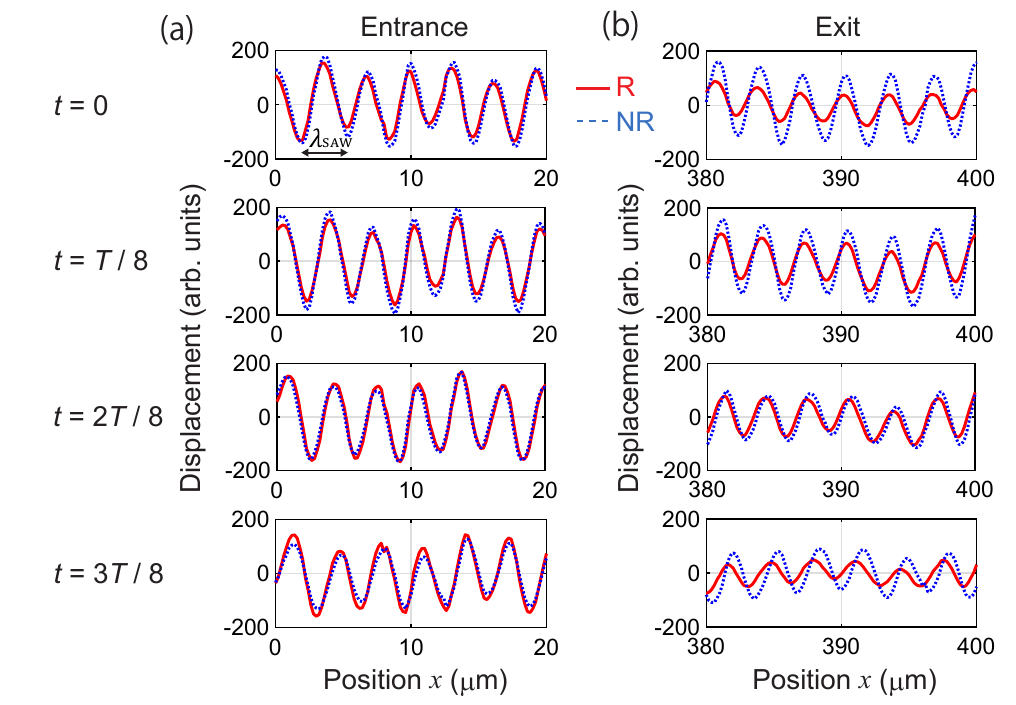}
    \caption{SAW-induced surface displacement profiles of the Ni thin film (a) near the entrance ($x = \SI{0}-\SI{20}{\micro m}$) and (b) near the exit ($x = \SI{380}-\SI{400} {\micro m}$) at different times $t$ ($t=0, \, T/8, \, 2T/8, \, 3T/8$, where $T$ represents the SAW oscillation period). The time origin is defined as the time instant when the surface displacement at $x=\SI{0}{\micro m}$ reaches its peak. The solid red curves are the profiles for resonant excitation ($\mu_0 H_{\text{\text{ext}}}$ = 3.3 mT), while the dotted blue curves are the those for $\mu_0 H_{\text{\text{ext}}}$ = 17.4 mT.}
     \label{time_displacement}
\end{figure*}

\subsection{Spatiotemporal dynamics of a SAW coupled to magnons}
To capture all details of the spatiotemporal dynamics of the SAW propagating on the Ni thin film, we performed pulsed laser interferometry. By recording both amplitude and phase of the SAW at each $x$ position, it is possible to trace the temporal evolution of the whole oscillation profile. Figure \ref{time_displacement}(a) shows the surface displacement profiles of the Ni thin film in the region near the IDT for SAW generation at different times $t$. This region of the Ni thin film is hereafter referred to as the entrance [Fig.~\ref{sample}; $x = \SI{0}-\SI{20}{\micro m}$]. Each panel within this figure shows the waveform of the SAW at a different time $t$ in terms of the SAW oscillation period $T$. Figure \ref{time_displacement}(b) shows the corresponding data at the exit [Fig.~\ref{sample}; $x = \SI{380}-\SI{400}{\micro m}$]. Furthermore, we compare the oscillation profiles obtained using the resonant excitation condition ($\mu_0 H_{\text{\text{ext}}} = \mu_0 H_{\text{\text{R}}}$ = 3.3 mT) and the non-resonant excitation condition ($\mu_0 H_{\text{\text{ext}}} = \mu_0 H_{\text{\text{NR}}}$ = 17.4 mT). These waveforms were reconstructed from the position-dependent amplitude and phase information of the interferometric signal, $R(x,\, \mu_0 H_{\text{ext}})$ and $\theta (x,\,\mu_0 H_{\text{ext}})$, respectively. All waveforms in Fig.~\ref{time_displacement} exhibit an oscillation with a wavelength of $\lambda_{\mathrm{SAW}}=\SI{3.2}{\micro m}$, and the SAW propagates in the $+x$ direction as time progresses.  At the entrance, the surface displacement profiles for resonant and non-resonant excitation are nearly identical. In contrast, at the exit, they differ significantly; compared to the non-resonant excitation condition, the data obtained using the resonant excitation condition has a smaller amplitude and a delayed phase. This finding implies  that a part of the energy of the surface vibration -- the phonon -- is transformed into magnon energy as the SAW propagates on the Ni thin film, because the effect of magnon--phonon coupling is stronger under resonant excitation conditions.

   To gain a deeper understanding of the local absorption of the surface vibrations in this sample, we compare the amplitude- and phase-profiles of the surface vibration in the resonant case with the data for non-resonant excitation. Figures \ref{norm_amp}(a) and \ref{norm_amp}(b) display the amplitude ratio $R(x,\,\mu_0 H_{\text{R}})/R(x,\, \mu_0 H_{\text{NR}})$ and the relative phase  $\theta (x,\,\mu_0 H_{\text{R}}) - \theta (x,\,\mu_0 H_{\text{NR}})$, respectively, as a function of the position $x$. The insets in these figures show magnified views around the entrance and exit. A distinct feature of these data is the presence of spatial oscillations, particularly at the entrance as shown in the left insets. Furthermore, it is clearly observed that the amplitude ratio decreases and the relative phase  becomes more negative as the SAW propagates on the Ni thin film. Finally, at the exit, these values become $\approx$ 0.6 and $\approx  -0.5$ rad, respectively. These values are close to the values determined by the RF transmission measurements (0.53 and $-0.51$ rad). These results verify that pulsed laser interferometry can quantitatively reproduce the SAW absorption at the exit evaluated by the conventional method (by measuring the $S_{21}$ parameter). Hence, pulsed laser interferometry is considered useful for the quantitative assessment of both the absorption and phase shift of a SAW across the entire sample.

\subsection{Characterization of the interference effect}

\begin{figure*}[!t]
    \centering
    \includegraphics[width=\textwidth]{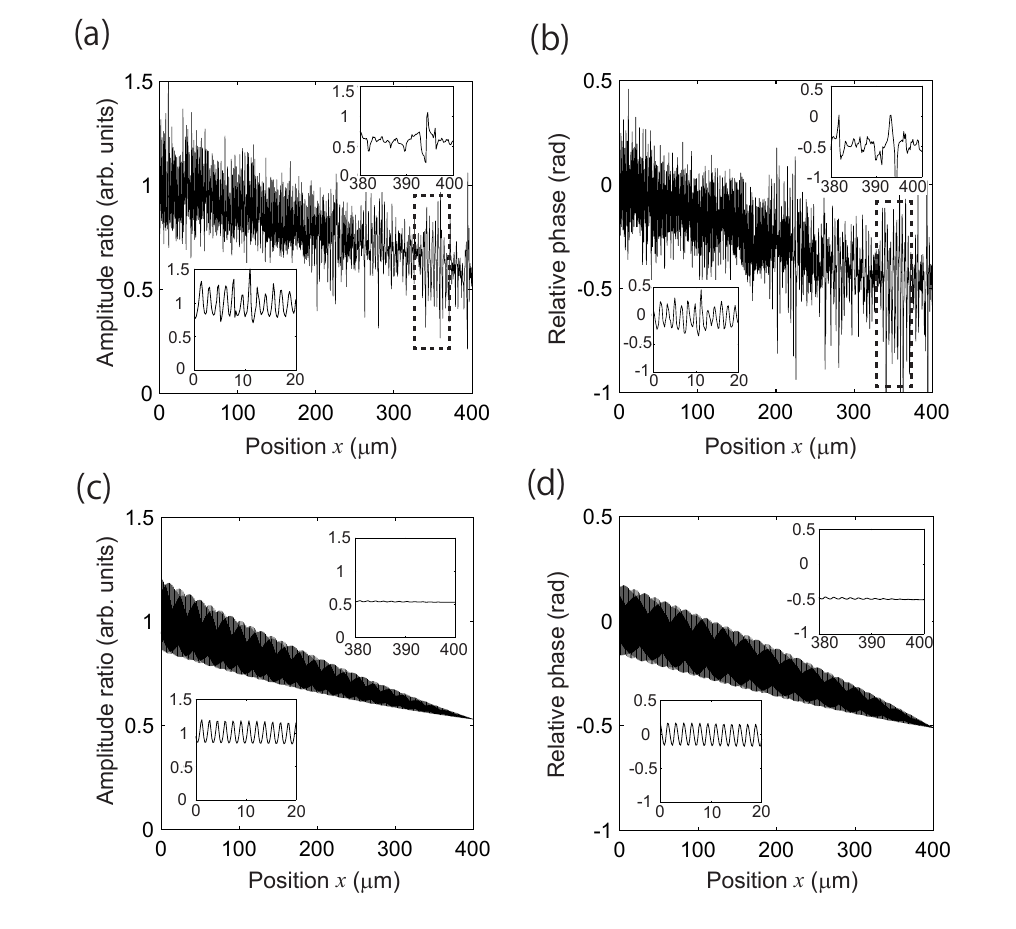}
    \caption{(a) The experimentally determined ratio of the surface displacement amplitude for resonant excitation to that for non-resonant excitation, $R(x,\mu_0 H_{\text{R}})/R(x,\mu_0 H_{\text{NR}})$, and (b) the experimentally determined relative phase  of the data for resonant excitation with respect to the phase for non-resonant excitation, $\theta (x, \mu_0 H_{\text{R}}) - \theta (x, \mu_0 H_{\text{NR}})$ as a function of $x$. (c) The analytically derived amplitude ratio and (d) the corresponding relative phase. Inside the dashed boxes in (a) and (b), additional oscillations can be confirmed, which are discussed in Section III.D. The insets in each graph provide magnified views of the data at the entrance and the exit. The $x$- and $y$-axes of the insets are in the same units as those of the main graph.
 }
     \label{norm_amp}
\end{figure*}
\begin{figure*}[!t]
    \centering
    \includegraphics[width=\textwidth]{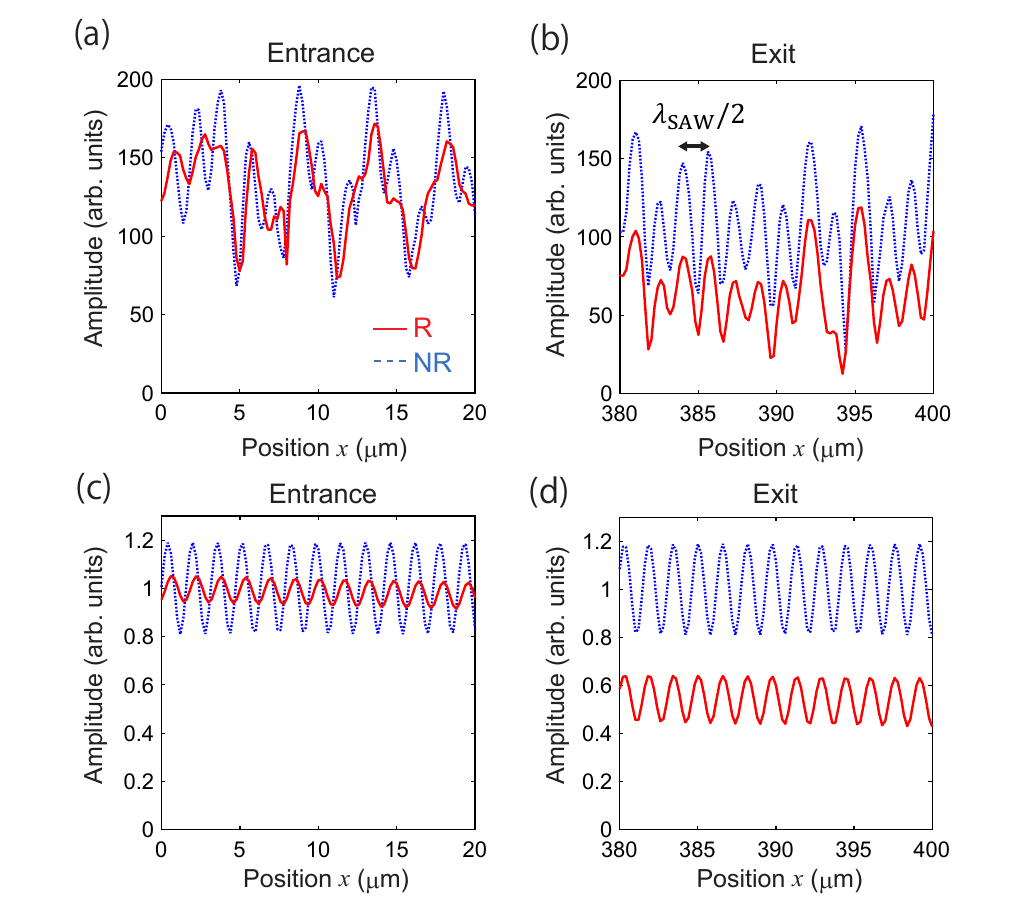}
    \caption{The experimentally determined spatial variations in the amplitude of the surface displacement near (a) the entrance and (b) the exit. The corresponding theoretically predicted spatial variations are shown in (c) and (d), respectively. The solid red curves are the amplitudes for  resonant excitation, and the dotted blue curves are those for non-resonant excitation.}
     \label{amplitude}
\end{figure*} \par

A distinct feature in Figs. \ref{norm_amp}(a) and \ref{norm_amp}(b) is the spatial oscillations, particularly prominent at the entrance position. Figures \ref{norm_amp}(c) and \ref{norm_amp}(d) show the theoretical predictions of the amplitude ratio and the relative phase. The detailed behavior is discussed later, but the most important point is that the oscillations in the insets are in good agreement with the experimentally observed oscillations.    To reveal the origin of the oscillations of the amplitude ratio in Fig.~\ref{norm_amp}(a), we analyze the behaviors of  the amplitude data near the entrance and the exit in Figs.~\ref{amplitude}(a) and \ref{amplitude}(b), respectively. (The analysis of the oscillations in Fig.~\ref{norm_amp}(b) is discussed in Appendix C.) In each figure, we show both $R(x,\, \mu_0 H_{\text{R}})$ and $R(x,\, \mu_0 H_{\text{NR}})$. All data sets have an oscillating component  with a wavelength of approximately $\SI{1.6}{\micro m}$, which is close to $\lambda_{\mathrm{SAW}}/2$   (the fluctuation of the amplitude that leads to peaks, shoulders, and valleys can be attributed to a bulk acoustic wave (BAW) contribution which will be discussed later in Section III.D). This suggests an interference effect due to multiple reflections of the SAW at the IDTs, which results in SAWs traveling in opposite directions. The obtained surface profiles cannot be explained by considering only a single SAW, because a single traveling SAW should result in a nearly constant amplitude (with a slight decay due to absorption).

   To reproduce the oscillating behavior observed in Figs.~\ref{norm_amp} and \ref{amplitude}, we developed an analytical model that describes the surface displacement of the Ni thin film caused by the SAW.  We first assume  that the surface displacement  is primarily influenced by two factors: the primary SAW that directly reaches the detection IDT, and the reflected SAW.  In this case, the ideal surface displacement under the non-resonant excitation condition, $u_{\text{NR}}$, can be written as: 

\begin{align}
    u_{\text{NR}}(x) &= A_{\text{NR}} \sin{(k_{\text{NR}} x-2 \pi f_{\text{SAW}} t+\phi_1)} \notag \\ 
     &+ B_{\text{NR}} \sin{(-k_{\text{NR}} x-2 \pi f_{\text{SAW}} t +\phi_1 + 2k_{\text{NR}} x_{0} + \Delta \theta)} 
     \label{u_disp}
\end{align}
and that under the resonant excitation condition becomes
\begin{align}
    u_{\text{R}}(x) &= A_{\text{R}}(x) \sin{(k_{\text{R}} x-2 \pi f_{\text{SAW}} t+\phi_1)} \notag \\ 
     &+ B_{\text{R}}(x) \sin{(-k_{\text{R}} x-2 \pi f_{\text{SAW}} t + \phi_1 + 2k_{\text{R}} x_{0} + \Delta \theta)}.
     \label{u_disp2}
\end{align}
Here, $A_{\text{NR}}$ and $B_{\text{NR}}$ are the SAW amplitudes of the primary and the reflected SAWs in the non-resonant case, and $k_{\text{NR}}$ is the wavenumber of the SAW in the non-resonant case. These amplitudes are considered to be constant due to the negligible SAW absorption in the ideal non-resonant case.  Furthermore, $B_{\text{NR}} = r_{\text{IDT}} A_{\text{NR}}$, where $r_{\text{IDT}}$ = 0.19 according to the RF transmission measurements. The phases $\phi_1$, $2k_{\text{NR}} x_{0}$ ($x_0$: length of the sample), and $\Delta \theta$ are the phases of the primary SAW at $x=0$ and $t=0$,   $2k_{\text{NR}} x_{0}$ is the phase offset of the reflected SAW at $x=0$ and $t=0$ in the case of propagation under the non-resonant excitation condition, and $\Delta \theta$ is the phase shift due to the round-trip between the edge of the Ni thin film and the IDT including the phase shift due to the reflection of the SAW at the detection IDT. The amplitudes of the primary and reflected SAWs in the resonant case depend on $x$ due to absorption (magnon--phonon coupling) \cite{Dreher2012}. We consider an exponential decay on the Ni thin film  with the absorption coefficient $\kappa$; $A_{\text{R}}(x)=A_{\text{NR}} e^{-\kappa x}$ and $B_{\text{R}}(x)=r_{\text{IDT}} A_{\text{NR}} e^{\kappa (x-2x_{0})}$. Here, we assumed that the amplitude and phase of the primary SAW at $x=0$ in Eq.~\eqref{u_disp2} are equal to those in Eq.~\eqref{u_disp}.  The values for $k_{\text{NR}}$, $k_{\text{R}}$, $\kappa$, $\phi_1$, and $\Delta \theta$ were determined from the pulsed laser interferometry and RF transmission measurements (further details are provided in Appendix \ref{parameter}).  
   From Eq.~\eqref{u_disp}, we obtain

\begin{align}
    R(x,\mu_0 H_{\text{NR}})=\sqrt{A_{\text{NR}}^2 + B_{\text{NR}}^2 + 2A_{\text{NR}}B_{\text{NR}} \cos{\bigl\{ 2k_{\text{NR}}(x_0-x)+\Delta \theta \bigr\}}}.
    \label{amp_NR}
\end{align}
The third term in the square root is the interference term, which produces nodes and antinodes in the spatial profile of the partial standing wave. Similarly, from Eq.~\eqref{u_disp2},we obtain
\begin{align}
    R(x,\mu_0 H_{\text{R}})=\sqrt{A_{\text{R}}(x)^2 + B_{\text{R}}(x)^2 + 2A_{\text{R}}(x)B_{\text{R}}(x) \cos{\bigl\{ 2k_{\text{R}}(x_0-x)+\Delta \theta \bigr\}}}.
    \label{amp_R}
\end{align}
$R(x,\mu_0 H_{\text{R}})$ also has nodes and antinodes, while its oscillation center, i.e. $A_{\text{R}}(x)$ as well as the oscillation amplitude, i.e. $B_{\text{R}}(x)$, exponentially decrease. Figures \ref{amplitude}(c) and \ref{amplitude}(d) show the spatial variations in the amplitude of the ideal surface displacement at the entrance and the exit, respectively, calculated using Eqs.~\eqref{amp_NR} and \eqref{amp_R}. At the entrance, the center of the amplitude oscillation is independent of the excitation condition, because $A_{\text{R}}(x) \approx A_{\text{NR}}(x)$ near $x$ = 0, while the oscillation amplitude depends on the excitation condition, because $B_{\text{R}}(x) < B_{\text{NR}}(x)$ near $x$ = 0. Furthermore, the difference between the positions of the antinodes of the SAWs for resonant and non-resonant excitation is largest at the entrance, because here the wavenumbers $k_{\text{NR}}$ and $k_{\text{R}}$ are different and $(x_0-x)$ is largest. These differences in the amplitude and the antinode position cause the spatial oscillation of the amplitude ratio shown in Fig.~\ref{norm_amp}(a). On the other hand, at the exit, the positions of the nodes and antinodes are similar, because here Eqs.~\eqref{amp_NR} and \eqref{amp_R} have similar phase values in their interference terms.

   Now we return to Fig.~\ref{norm_amp}(c), where we showed the theoretically predicted amplitude ratio calculated using  Eqs.~\eqref{amp_NR} and \eqref{amp_R}. At the entrance of the Ni thin film, the amplitude ratio oscillates with a wavelength of $\lambda_{\text{SAW}}/2$, while at the exit, the amplitude ratio is almost constant. The reason for  the nearly constant amplitude ratio at the exit is the combination of the above-mentioned similar phase and the similar reduction of the amplitudes of the primary SAW ($A_{\text{R}}(x) \approx A_{\text{NR}} e^{-\kappa x_0}$) and the reflected SAW ($B_{\text{R}}(x) \approx B_{\text{NR}} e^{-\kappa x_0}$) near the exit:

\begin{align}
    \frac{R(x\sim x_0,\mu_0 H_{\text{R}})}{R(x \sim x_0,\mu_0 H_{\text{NR}})}\approx e^{-\kappa x_0},
    \label{exit_amp_ratio}
\end{align}
In Fig.~\ref{norm_amp}(d), we showed the theoretically predicted relative phase  calculated from Eqs.~\eqref{u_disp} and \eqref{u_disp2}. The relative phase  also exhibits a pronounced oscillation at the entrance and is nearly constant at the exit. This behavior is discussed in Appendix \ref{ampphase}.

   Figure \ref{norm_amp} confirms that the overall behavior of the experimentally determined spatial profiles, including the local oscillation behavior, can be well explained by a simple analytical model. This agreement strongly suggests that the interference between the primary SAW and the reflected SAW governs the oscillating surface profile across the entire the sample. Thus, our results clearly visualize the long-range propagation of the SAW coupled to magnons, which travels over a distance on the order of millimeters (at least twice the length of the Ni thin film).

\subsection{Additional features}
\begin{figure*}[!t]
    \centering
    \includegraphics[width=\textwidth]{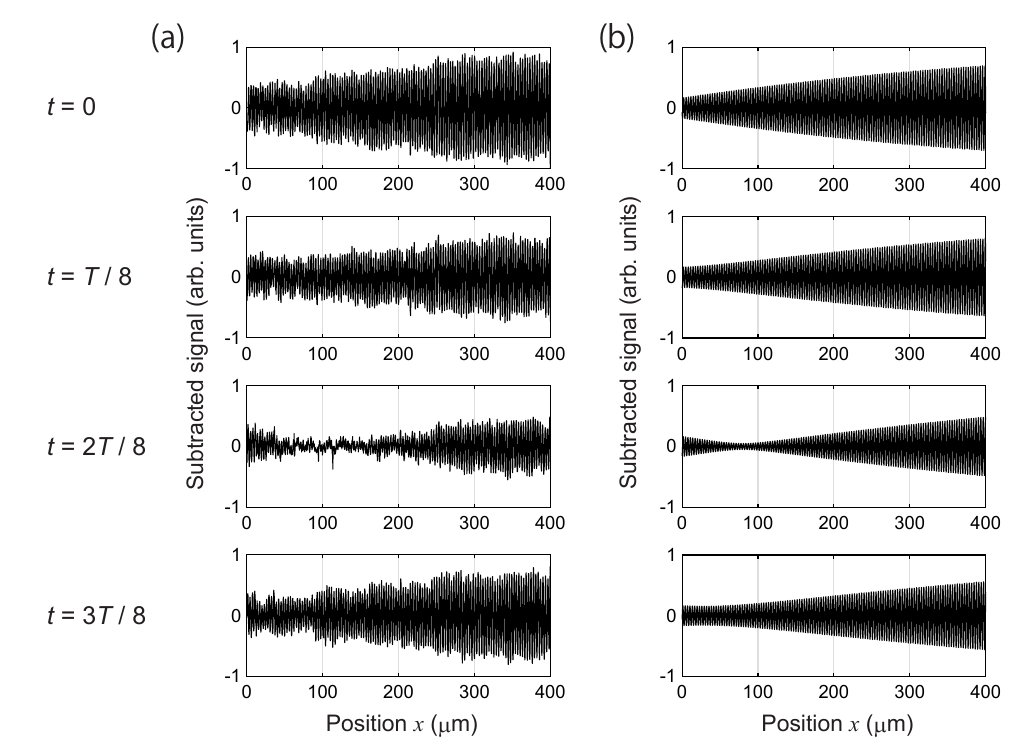}
    \caption{Difference between the surface profiles obtained using resonant and non-resonant excitation conditions for $t=0, \, T/8, \, 2T/8,$ and $3T/8$. (a) Experimental results. (b) Ideal result $u_{\text{R}}(x)-u_{\text{NR}}(x)$.}
     \label{dif_amp}
\end{figure*} \par
The experimental amplitude ratio and relative phase  presented in Figs.~\ref{norm_amp}(a) and \ref{norm_amp}(b) contain also features that cannot be explained by our simple model, particularly around $x =  \SI{360}{\micro m}$ [see the dashed boxes in Figs.~\ref{norm_amp}(a) and \ref{norm_amp}(b)]. We consider that this feature does not originate from the Rayleigh-type SAW propagation. A possible origin may be a bulk acoustic wave produced by the IDT \cite{Knuuttila2001,Lu2020,Lu2021}. Such a wave can be reflected at the bottom of the substrate, and thus can also influence the sample surface.  If this additional feature is independent of the external magnetic field, it can be removed by subtracting the displacement profile for resonant excitation from the displacement profile for non-resonant excitation. Figure \ref{dif_amp}(a) shows the subtracted waveforms derived from Fig.~\ref{time_displacement} for four specific times $t$. For comparison, Fig.~\ref{dif_amp}(b) shows the corresponding theoretical result, $u_{\text{R}}(x)-u_{\text{NR}}(x)$ for the same times $t$. The experimental and theoretical results show a slightly better agreement than in Fig. 4, which indicates that this subtraction method effectively cancels the influence of the additional features. This agreement supports our conclusion that the additional features are not related to magnon--phonon coupling.

    The period of the subtracted waveforms in Figs.~\ref{dif_amp}(a) and \ref{dif_amp}(b) is approximately $\lambda_{\text{SAW}}$, and these waveforms reflect the differences in the amplitude and phase of the surface displacements obtained using the resonant and non-resonant excitation conditions. The oscillation amplitude of the subtracted signal at the exit is larger than that at entrance due to the accumulated phase difference of the SAW, $(k_{\text{NR}}-k_{\text{R}} )x_0$. We also note that the oscillation amplitude at the entrance is significant. Because the vibration profile of the primary SAW at the entrance is almost independent of the excitation condition, the subtraction should not result in a significant amplitude oscillation if the primary SAW is dominant at the entrance. This implies that the interference effect is also important at the entrance.

\section{conclusion}
We have demonstrated that the spatiotemporal dynamics of a SAW coupled to magnons can be visualized across a 0.4-mm-long Ni thin film by pulsed laser interferometry. Both a decrease in the amplitude and an increasing absolute value of the relative phase  were observed as the SAW propagates. These experimental results are in agreement with the results of our RF transmission measurements.  We have shown that the amplitude ratio and relative phase  due to the magnon--phonon coupling can be reproduced by a simple analytical model  that accounts for the absorption, phase delay, and reflection of the SAW in the sample.  Our data clarifies that the SAW coupled to magnons  travels over a distance on the order of millimeters due to reflection, and this significantly affects the surface vibration profile. Our findings provide a clear interpretation of the surface profile of a system with magnon--phonon coupling induced by long-range SAW propagation.

   We consider that this interferometric method enables us to understand magnon--phonon coupling in spatially inhomogeneous samples, where the absorption profiles are non-uniform. For example, this method can facilitate the visualization of surface vibrations coupled to magnons in various phononic structures \cite{Wang2022}. Furthermore, various SAW-induced phenomena \cite{Nie2023} can be traced by observing changes in the surface profile. Thus, this research may help to unveil novel SAW-induced phenomena.

\begin{acknowledgments}
This work was partially supported by JST CREST (Grant No. JPMJCR19J4), JSPS KAKENHI (Grant No. JP23H01471), the MEXT Quantum Leap Flagship Program  (MEXT Q-LEAP) (Grant No. JPMXS0118067246), and the Precise Measurement Technology Promotion Foundation (PMTP-F). Part of this work  was conducted at the AIST Nano-Processing Facility supported by the "Nanotechnology Platform Program" of the Ministry of Education, Culture, Sports, Science and Technology (MEXT), Japan (Grant No. JPMXP09-F-21-AT-0085).
\end{acknowledgments}

\appendix

\section{Details of the pulsed laser interferometry measurements}
In this Appendix, we provide further details of the measurement procedure. First, the output of the mode-locked Ti:Sapphire laser  with a center wavelength of 806 nm was optimized with respect to power and polarization by a half-wave plate and a polarizer  so that the two beams entering the BP in Fig.~\ref{sample}(c) from Path 1 and Path 2 have the same intensity. Then, the pulses were introduced into the Michelson-interferometer unit. Within this unit, the laser beam was split into Path 1 (S-polarized) and Path 2 (P-polarized) using the PBS. Here, S-polarization means that the direction of the electric field is perpendicular to the plane of incidence and P-polarization means that the direction of the electric field is parallel to the plane of incidence. The S-polarized light in Path 1 traveled through a quarter-wave plate (QWP), passed through the objective lens,  and after the reflection at the sample it passed through the objective lens and the QWP again, which results in P-polarized light that can pass through the PBS. Similarly, the originally P-polarized pulses in Path 2 became S-polarized after reflection at the mirror on the translation stage in Path 2 and passing through the QWP again. These pulses were reflected by the PBS. The two beams were thus combined at the PBS, and then the interferometric signal was measured by the upper photodiode  of the BP in Fig.~\ref{sample}(c).

   Note that a fraction of the beam in Path2 was reflected toward the other photodiode by a variable neutral density filter in order to cancel the dc component of the interference signal. Furthermore, the mirror on the translation stage was mounted on a piezoelectric actuator for fine adjustments of the optical path length (the translation stage was used for coarser adjustments). The feedback control circuit for the piezoelectric actuator had a bandwidth of $\approx$ $\SI{1}{kHz}$ and ensured an almost constant difference between the optical path lengths. The modulation frequency of the interferometric signal ($\SI{70}{kHz}$) was much higher than the bandwidth of the feedback control circuit  to ensure that the signal is not influenced by the feedback. The implemented feedback control was able to effectively reduce  the impact of small vibrations  of the optical components. This experimental approach is based on the work by Shao {\it et al.} \cite{Shao2022}. The key difference between our experiment and this previous work is that all RF frequencies ($f_{\text{SAW}}$ and reference frequency used for lock-in detection) are  referenced to the laser pulse repetition rate in our work. Therefore, the modulation frequency used for lock-in detection remained stable.

\section{Determination  of the input parameters for the analytical model}
\label{parameter}
The wavenumber $k_{\text{NR}}$ was determined by calculating the Fourier transform of the observed surface vibration profile for non-resonant excitation, and we obtained $k_{\text{NR}} = 2 \pi \times  \SI{0.3172}{\micro m}^{-1}$. Then, to reproduce the experimental results shown in Fig.~\ref{dif_amp}(a), we set the initial phase of the primary SAW to $\phi_1=0.44 \pi$. The value of $\Delta \theta$ was  chosen in such a way to reproduce the position of the local maximum at $x=\SI{0.4}{\micro m}$ for non-resonant excitation in Fig.~\ref{amplitude}(a).

    The wavenumber $k_{\text{R}}$ can be inferred from $ k_{\text{NR}}$ and the phase delay $\phi_{\text{R}} \equiv - \text{arg}(S_{\text{21}}(\nu_0, \mu_0 H_{\text{R}}))_{\text{norm}}$ = 0.51 observed in the RF transmission measurements in Fig.~\ref{VNA}(c):
\begin{align}
    k_{\text{R}} = k_{\text{NR}}  \cdot \frac{1}{1-\frac{1}{n_{\text{NR}}} \cdot \frac{\phi_{\text{R}}}{2\pi}}.
\label{disp_k_R}
\end{align}
Here, $n_{\text{NR}}=k_{\text{NR}} x_0 /2\pi=126.88$ is the number of cycles of the SAW on the Ni thin film under the non-resonant excitation condition. From Eq.~\eqref{disp_k_R}, we obtained $k_{\text{R}} = 2 \pi \times \SI{0.3174}{\micro m}^{-1}$. The damping coefficient was calculated by considering the decay over the distance $x_0$:  $\kappa = -\log (|S_{21}(\nu_0, \mu_0 H_{\text{R}})|_{\text{norm}})/ x_0 =1.59\times 10^{-3} \, \si{\micro m}^{-1}$, using $|S_{21}(\nu_0, \mu_0 H_{\text{R}})|_{\text{norm}}=0.53$ shown in Fig.~\ref{VNA}(b).

\section{Spatial variation of the relative phase}
\label{ampphase}
 
In this Appendix, we analytically discuss the spatial variation of the relative phase  shown in Fig.~\ref{norm_amp}(d). From Eq.~\eqref{u_disp}, we can derive a spatially oscillating phase for $t=0$ due to interference:
\begin{align}
    \theta (x, \mu_0 H_{\text{NR}})=-k_{\text{NR}}x_0 -\phi_1 -\delta_{\text{NR}}(x)+\pi,
    \label{phase_NR}
\end{align}
where $\delta_{\text{NR}}(x)$ represents the polar angle of the point,
\begin{align}    
\delta_{\text{NR}}(x) \colon \Bigl( &A_{\text{NR}} \cos{\{k_{\text{NR}}(x_0 -x)\}}+B_{\text{NR}} \cos{\{k_{\text{NR}}(x_0 -x) + \Delta \theta\}} \, , \notag\\
  -&A_{\text{NR}} \sin{\{k_{\text{NR}}(x_0 -x)\}}+B_{\text{NR}} \sin{\{k_{\text{NR}} (x_0 -x) + \Delta \theta\}} \Bigr).
  \label{delta_NR}
\end{align}
Similarly, we can also derive a spatially oscillating phase from Eq.~\eqref{u_disp2}:
\begin{align}
    \theta (x, \mu_0 H_{\text{R}})=-k_{\text{R}}x_0 -\phi_1 -\delta_{\text{R}}(x)+\pi,
    \label{phase_R}
\end{align}
where $\delta_{\text{R}}(x)$ represents the polar angle of the point,
\begin{align}    
\delta_{\text{R}}(x) \colon \Bigl( &A_{\text{R}} (x) \cos{\{k_{\text{R}}(x_0 -x)\}}+ B_{\text{R}} (x) \cos{\{k_{\text{R}}(x_0 -x) + \Delta \theta\}} \, , \notag\\
  -&A_{\text{R}} (x) \sin{\{k_{\text{R}}(x_0 -x)\}}+B_{\text{R}} (x) \sin{\{k_{\text{R}} (x_0 -x) + \Delta \theta\}} \Bigr). 
  \label{delta_R}
\end{align}
From Eqs.~\eqref{phase_NR} and \eqref{phase_R}, we obtain the analytical expression for the relative phase: 
\begin{align}
        \theta (x, \mu_0 H_{\text{R}})-\theta (x, \mu_0 H_{\text{NR}})=(k_{\text{NR}}-k_{\text{R}})x_0 +\delta_{\text{NR}}(x)-\delta_{\text{R}}(x).
    \label{phase_diff_calc}
\end{align}

This equation can be used to discuss the oscillation profile shown in Fig.~\ref{norm_amp}(d): Firstly, we consider the spatial variation of the relative phase  near the exit ($x \approx x_0$). Here, the relationships $A_{\text{R}} \approx A_{\text{NR}}e^{-\kappa x_0}$ and $B_{\text{R}} \approx B_{\text{NR}}e^{-\kappa x_0}$ hold. Furthermore, also the phases in the cosine and sine functions in Eqs.~\eqref{delta_NR} and \eqref{delta_R} are similar since $(x_0-x) \approx 0$, and thus we have $\delta_{\text{NR}}(x) \approx \delta_{\text{R}}(x)$. Consequently, the relative phase  near the exit can be approximately described by
\begin{align}    
\theta (x \sim x_0, \mu_0 H_{\text{R}})-\theta (x \sim x_0, \mu_0 H_{\text{NR}}) \approx (k_{\text{NR}}-k_{\text{R}})x_0,
  \label{exit_phase_shift}
\end{align}
which is independent of the position. As a result, the relative phase  near the exit in Fig.~\ref{norm_amp}(d) is almost constant.

   Secondly, we consider the relative phase  close to the entrance ($x \approx 0$). In this region, the position-dependent term of the relative phase, $\delta_{\text{NR}}(x) - \delta_{\text{R}}(x)$, plays an important role in shaping the profile of the relative phase, because the amplitude of the reflected SAW under the resonant excitation condition is relatively small ($B_{\text{R}} (x) < B_{\text{NR}}$), whereas the amplitude of the primary SAW is not strongly affected by the excitation condition ($A_{\text{NR}} (x) \approx A_{\text{R}}$). Furthermore, the difference between the phases defined in Eqs.~\eqref{delta_NR} and \eqref{delta_R} is largest at the entrance. These properties determine the spatial profile of the relative phase of the interfering SAWs.

\bibliography{apssamp}

\end{document}